Research article

# Metal-insulator transition and electroresistance in lanthanum/calcium manganites La$_{1-x}$Ca$_x$MnO$_3$ ($x$ = 0–0.5) from voltage-current-temperature surfaces

JC Knott, DC Pond and RA Lewis*

Address: Institute for Superconducting and Electronic Materials, University of Wollongong, Wollongong NSW 2522, Australia

Email: JC Knott - jck422@uow.edu.au; DC Pond - dcp20@uow.edu.au; RA Lewis* - roger@uow.edu.au

* Corresponding author

## Abstract

Of the perovskites, ABX$_3$, a subset of special interest is the family in which the A site is occupied by a lanthanide ion, the B site by a transition metal and X is oxygen, as such materials often exhibit a large change in electrical resistance in a magnetic field, a phenomenon known as "colossal" magnetoresistance (MR). Two additional phenomena in this family have also drawn attention: the metal-insulator transition (MIT) and electroresistance (ER). The MIT is revealed by measuring resistance as a function of temperature, and observing a change in the sign of the gradient. ER – the dependence of the resistance on applied current – is revealed by measuring resistance as a function of applied current. Up until now, the phenomena of MIT and ER have been treated separately. Here we report simultaneous observation of the MIT and ER in the lanthanum/calcium manganites. We accomplish this by measuring voltage-current curves over a wide temperature range (10–300 K) allowing us to build up an experimental voltage surface over current-temperature axes. These data directly lead to resistance surfaces. This approach provides additional insight into the phenomena of electrical transport in the lanthanum/calcium manganites, in particular the close connection of the maximum ER to the occurrence of the MIT in those cases of a paramagnetic insulator (PMI) to ferromagnetic metal (FMM) transition.

## Background

Electroresistance (ER) is a change in electrical resistance with applied voltage or applied current. A material that obeys Ohm's law has zero ER. So to say a material exhibits ER is equivalent to saying it is non-ohmic. A third way of saying the same thing is by referring to the voltage-current (V-I) characteristic. An ohmic material shows a linear relationship between voltage and current. In an electroresistive material the V-I relationship is non-linear.

The term "electroresistance" has sprung into vogue on the back of the cognate term "magnetoresistance", which refers to a change of electrical resistance with magnetic field. Magnetoresistance (MR) is defined by

$$\text{MR} = \frac{R(B)\text{-}R(0)}{R(0)} = \frac{\Delta R}{R},\tag{1}$$

where $R(B)$ is the electrical resistance measured in applied magnetic field $B$. MR has broad implications both for fundamental physics and for technological applications, the latter particularly in relation to memory devices. Substantial magnitudes of MR have lead to the coining of the terms "giant magnetoresistance" (GMR), now mainly associated with metallic layer systems, and "colossal magnetoresistance" (CMR), now mainly associated with perovskite oxides. As the magnetic field typically reduces the electrical resistance, MR as defined in Eqn. 1 is typically negative; for this reason the opposite definition is sometimes employed. Sometimes the resistance at field is used as the denominator, usually with the effect of making the magnitude of the ratio larger than one (sometimes, much larger).

In analogy to Eqn. 1, ER may be defined in terms of the resistance at the applied electric field $E$,

$$\text{ER}(E,0) = \frac{R(E)-R(0)}{R(0)} = \frac{\Delta R}{R}.\tag{2}$$

This formulation is appropriate for experiments conducted in a field-effect (FE) configuration, in which the applied bias is provided by a gate, separate to the voltage drop between source and drain used to measure the sample resistance [1,2]. A more common arrangement is to use the standard four-probe configuration to measure the sample resistance. The current is applied across two outer terminals and the voltage measured between two inner terminals. In this configuration it is not possible to obtain data at zero current so the resistance at a reference current is employed in the definition of ER. Consider measurements made at the two currents $I_{low}$ and $I_{high}$. Each of four separate combinations to define electroresistance have been used in the literature.

$$\text{ER} = \frac{R(I_{high})-R(I_{low})}{R(I_{low})}\tag{3}$$

is used by Refs. [3-6]. Refs. [7,8] use the opposite ("ER-bar"):

$$\overline{\text{ER}} = \frac{R(I_{low})-R(I_{high})}{R(I_{low})}.\tag{4}$$

Alternatively, Refs. [9,10] use

$$\mathrm{ER}^* = \frac{R(I_{\mathrm{low}}) - R(I_{\mathrm{high}})}{R(I_{\mathrm{high}})}, \tag{5}$$

("ER-star") while Refs. [11,12] use

$$\overline{\mathrm{ER}}^* = \frac{R(I_{\mathrm{high}}) - R(I_{\mathrm{low}})}{R(I_{\mathrm{high}})}, \tag{6}$$

("ER-star-bar"), and other definitions still are possible based on the maximum resistance [13] or dV/dI [14]. Often the definition chosen is the one in which the result turns out to be large and positive. The measurements we report here are taken using the four-terminal method. We will use Eqn. 3 to define ER, as that seems most closely connected to the field-effect configuration definition, Eqn. 2, and the definition of magnetoresistance, Eqn. 1. As with magnetoresistance, the more emphatic terms "giant" ER (GER) [4,6,14-16] and "colossal" ER (CER) [2,3,7,8,10-13,17-22] have been adopted, although no specific criteria appear to have been articulated as to when ER becomes GER or CER.

A prosaic origin of ER is joule heating. Consider a typical metal. Its resistance increases with temperature. If a large bias is applied, a large current will flow, leading to a large amount of joule heating, leading to an increase in temperature, leading to increase in resistance, and hence positive ER. On the other hand, a semiconductor, having a negative temperature coefficient of electrical resistance, will show negative ER as a consequence of joule heating. While joule heating is a legitimate way to induce ER, and the basis of the technical application of the ER effect [23], it is generally to be avoided where more subtle mechanisms of ER are to be identified [24,25]. This is the case in the present experiments.

In the last paragraph it was noted that electrical conductors and insulators may be differentiated depending on whether the temperature coefficient of resistance is positive (conductors) or negative (insulators). This is a fundamental way to distinguish whether a material is a conductor or an insulator – rather than on the basis of the magnitude of the resistivity – and the criteria used, for example, in attempting to determine if the ground state of a two-dimensional electron gas is metallic or not. It turns out that many materials that exhibit MR, and ER, undergo a change on increasing temperature from a metallic state to an insulating state, evidenced by a change in the sign of $dR/dT$. This is a temperature-induced metal-insulator transition (MIT). The experiments we report here detect the MIT through presenting resistance as a function of temperature.

ER has been reported in a variety of distinct physical systems. (a) Non-linear I-V characteristics are commonplace in electronic systems based on semiconductors, either conventional or low-dimensional, but are not generally referred to in terms of ER. (b) ER has been described in metal-ferroelectric-metal junctions (ferroelectric tunnel junctions), such as Pt/BaTiO$_3$/Pt [26], and has as its origin direct quantum tunnelling [15]. (c) ER has been obtained above a temperature-dependent threshold in amorphous carbon film on a silicon substrate [13]. (d) A strong ER, which is well-suited to switching operations, has been found in systems involving a metal electrode or electrodes and an oxide perovskite. Various metals (Pt, Au, Ag, Al, Ti, Mg) in conjunction with La$_{1-x}$Sr$_{1+x}$MnO$_4$ crystals give the maximum effect for $x \sim$0.5 [21]. SrRuO$_3$/SrTi$_{1-x}$Nb$_x$O$_3$ Schottky junctions also have useful switching characteristics [20], as do films of Pr$_{0.63}$Ca$_{0.37}$MnO$_3$ [27] and Pr$_{0.7}$Ca$_{0.3}$MnO$_3$ between metallic electrodes [3,19]; the proposed mechanism is a Mott transition [17] or electrochemical migration [28].

Much work on ER has been reported for thin films and single crystals of manganites. The most extensively-studied composition is La$_{0.7}$Ca$_{0.3}$MnO$_3$ [1,2,4,5,25,29]. This composition is the member of the La$_{1-x}$Ca$_x$MnO$_3$ series with the highest temperature of the MIT [30]; it changes from paramagnetic insulator (PMI) to ferromagnetic metal (FMM) at about 250 K when in the bulk form. ER in this compound is generally attributed to a percolative phase separation (PS) [1]. Formed into a *p-n* junction with a suitable substrate, a diodic behaviour is observed [12,31]. The ER has been attributed to current shunting [12]. A *p-i-n* diode structure has also been fabricated [32]. The composition La$_{0.8}$Ca$_{0.2}$MnO$_3$ shows similar behaviour [33], with the mechanism again taken to be related to PS [18,34]. ER has also been observed in the related compounds La$_{0.7}$Ce$_{0.3}$MnO$_3$ [10], which is electron doped, La$_{0.67}$Sr$_{0.33}$MnO$_3$ [6,9,14], Nd$_{0.65}$Ca$_{0.35}$MnO$_3$ [23], La$_{0.9}$Ba$_{0.1}$MnO$_3$ [35], $R_{0.67}$Ca$_x$MnO$_3$ [24], Ca$_{0.9}$Ce$_{0.1}$MnO$_3$ [16], as well as the bi-layered manganite La$_{1.2}$Sr$_{1.8}$Mn$_2$O$_7$ [11]. As the fraction $x$ in La$_{1-x}$Ca$_x$MnO$_3$ decreases below $\sim$0.2 the low-temperature state is a ferromagnetic insulator (FMI). In La$_{0.82}$Ca$_{0.18}$MnO$_3$ [8] and other FMI materials Nd$_{0.7}$Pr$_{0.3}$MnO$_3$ [7] and La$_{0.9}$Sr$_{0.1}$MnO$_3$ [22,33,36] it is observed that the ER and MR effects are decoupled. In ultrathin films [25] the non-linear characteristics are not intrinsic; extrinsic factors include contact resistance and joule heating. effects observed in thicker films are thought to arise due to grain structure and disorder [4].

The aim of the research to be reported here is to give a comprehensive experimental description of ER and MIT in lanthanum/calcium manganites by measuring V-I curves at many temperatures and so present the voltage (and hence resistance) data as a function of both current and temperature. Three aspects distinguish this work from previous studies. First, the work deals exclusively with bulk materials. Most previous studies have concentrated solely on thin films, in which the ER (and MR) effects are more pronounced. The present work is more fundamental in that the complications of lattice strain and anisotropy do not enter into the interpretation of the results. Second, this work is a systematic study of the series La$_{1-x}$Ca$_x$MnO$_3$ for $x$ = 0–0.5. Many previous reports on ER concentrate on a single composition. The wider view here gives a more complete picture of the ER phenomenon. Third, the work reported here is comprehensive in that a large number of V-I curves are presented at many temperatures; put another way, a large number of R(esistance)-T sets are presented at different currents. To give an idea of the scale of the difference, an earlier reports gives V-I curves at 6 temperatures [11]; we use dozens of temperatures. Other reports give R-T curves at 3 or 6 voltages [1,2] or 3, 4 or 12 currents [4,9,10]; we use hundreds of currents. In short, others give curves, we give surfaces. This larger data set assists in visualization, and confirms which behaviour is systematic and does not depend on specific choice of experimental conditions.

## Experimental details

The samples were synthesized by a partial melting technique as described previously [37]. High-purity starting materials were weighed, mixed, pelletised and sintered at 1100°C for 24 hours then ground, pelletised, and partially melted at 1500°C for 10 minutes before rapid cooling to 900°C. X-ray diffraction peaks were well-indexed by a rhombohedral structure. Only a single phase was evident. We did not specifically examine oxygen deficiency in the samples but the systematic data to be presented below suggests it does not play an important role in these measurements. Variable temperature measurements were made using a Janis CCS350R closed-cycle cooler. Electrical contacts were made using Ag paint. The current was supplied and the voltage measured with a Keithley SourceMeter 2400.

## Results and discussion

Our experimental results appear in Figs. 1, 2, 3, 4, 5, 6, 7, 8, 9, 10, 11, 12, which give voltage and resistance surfaces for $La_{1-x}Ca_xMnO_3$ for $x$ = 0, 0.1, 0.2, 0.3, 0.4 and 0.5. In the voltage (odd-numbered) figures, the voltage drop across the sample is shown on the vertical axis as a function of the temperature of the sample and the current applied to the sample. The vertical panes perpendicular to the current axis give the temperature dependence of the voltage, or, within a multiplicative factor, the temperature dependence of the resistance or of the resistivity. The temperature-induced MIT may be observed with reference to data in these planes. The vertical planes perpendicular to the temperature axis give the V-I characteristic. If the cross-section of the data with one of these planes is not a straight line, then the sample exhibits ER. In the resistance (even-numbered) figures, the resistance of the sample (simply obtained by dividing V by I; not the differential resistance, dV/dI) is shown on the vertical axis. The vertical panes perpendicular to the current axis give the R-T curves and directly allow observation of the MIT. The vertical planes perpendicular to the temperature axis give the R-I characteristic; if these are not cut horizontally by the data, the sample exhibits ER. A detailed discussion of each of the data sets is now given, with reference to the phase diagram of $La_{1-x}Ca_xMnO_3$ [30,38-40]. At room temperature, each sample is a PMI.

### x = 0

From Figs. 1 and 2, this compound appears to be insulating at all temperatures. The data on the planes perpendicular to the current axis in Fig. 2 are very similar to the resistivity-temperature data of Okuda *et al.* [41]. In that experiment, the data gives no evidence of a phase transition with temperature, although magnetization and other data suggest a phase transition at ~160 K to a state variously described as FMI [30], FM droplets or FM/AF coexistence [39], or canted antiferromagnetic (CAF) [40]. When examined closely, we observe a small bump in the R-T data at 185 K. This may correspond to the momentary appearance of a FMM phase, as is observed by others over a small temperature range in the $x$ = 0.18 compound [8]. In experiments that involve FMI materials, CER has been reported in the absence of MR [7,8]. The data on the planes perpendicular to the temperature axis in Fig. 2 is the R-I data. The surface cuts these planes almost horizontally, in other words, the resistance hardly varies with current. This demonstrates that there is very little ER in this material (estimated to be <10%). To our knowledge, ER or the lack of it has not been reported previously in a sample of this composition.

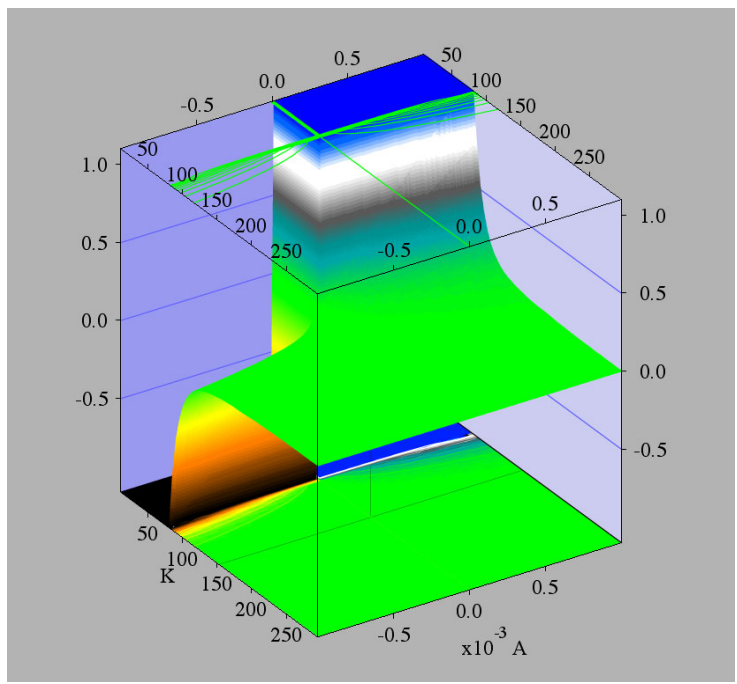

**Figure 1**
**LaMnO₃ voltage surface**. The voltage across the sample is shown on the vertical axis as a function of the applied current and the temperature.

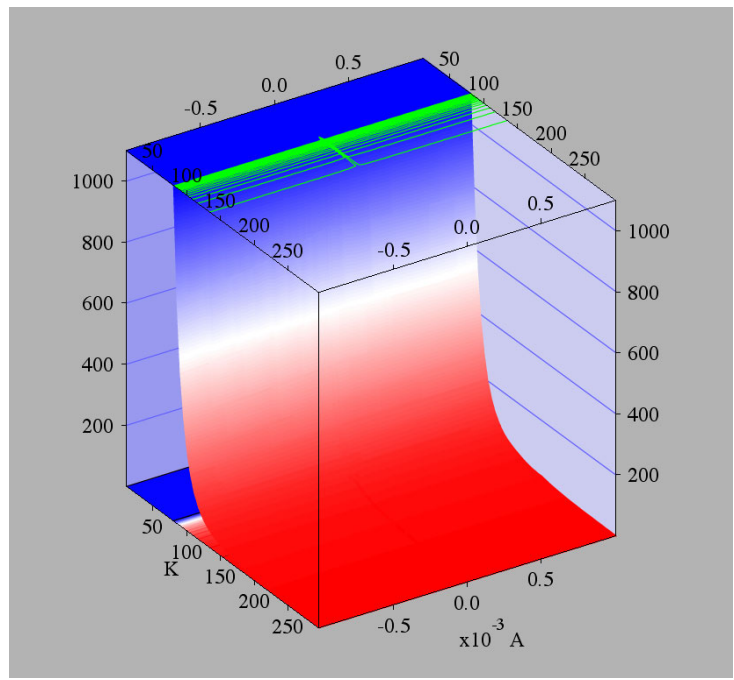

**Figure 2**
**LaMnO₃ resistance surface**. The resistance of the sample (Ω) is shown on the vertical axis as a function of the applied current and the temperature.

### *x* = 0.1

According to our data presented in Figs. 3 and 4, this sample undergoes a MIT at the temperature of the resistance peak, $T_p \sim$ 80 K. Previous measurements have only given data for $T >$ 100 K [42] and $T >$ 125 K [41]. The phase diagrams predict a transition to an FMI state [30], to FM droplets or FM/AF coexistence [39], or to an FMI then finally to a charge-ordered (CO) state [40]; we presume the change in gradient of the R-T data observed is related to the onset of the CO phase. As with the $x$ = 0 sample (Fig. 2), the very uniform value of resistance as the current is varied at a given temperature (Fig. 4), indicates there is very small ER (<10%) at all temperatures. To our knowledge, ER or the lack of it has not been reported previously in a sample of this composition.

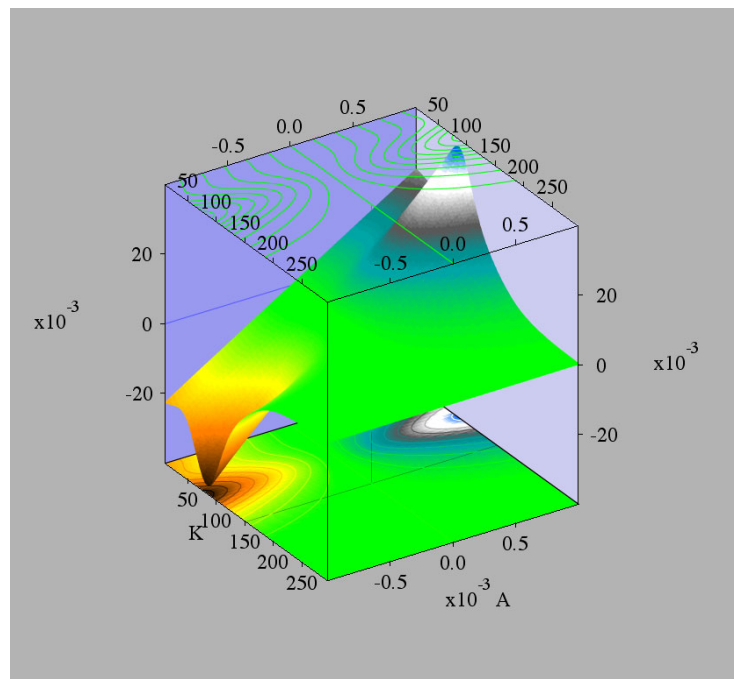

**Figure 3**
**La$_{0.9}$Ca$_{0.1}$MnO$_3$ voltage surface**. The voltage across the sample is shown on the vertical axis as a function of the applied current and the temperature.

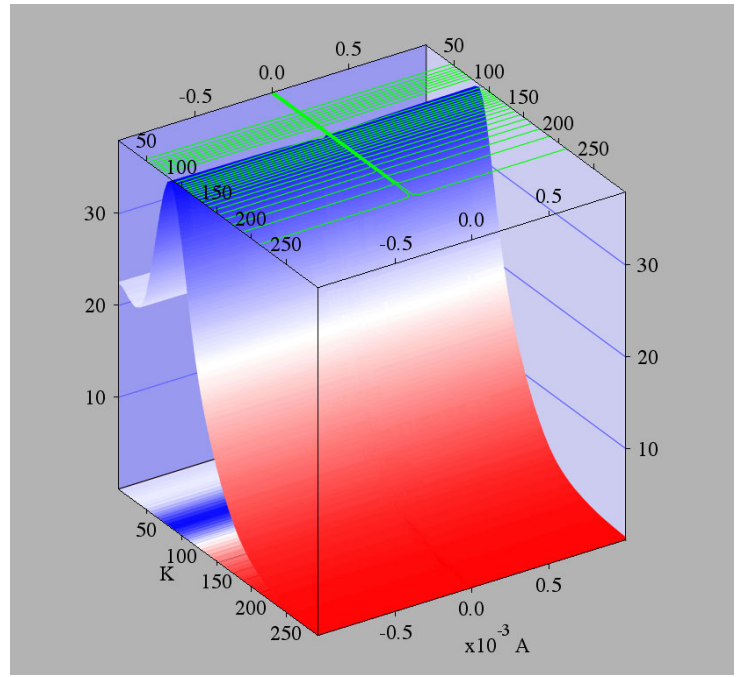

**Figure 4**
**La$_{0.9}$Ca$_{0.1}$MnO$_3$ resistance surface**. The resistance of the sample ($\Omega$) is shown on the vertical axis as a function of the applied current and the temperature.

---

### *x = 0.2*

At *x* = 0.18, just below the *x*-value of the sample to be next discussed, one report has been shown a closely-correlated ER and MR [43], with the ER being attributed to an electrically-induced MR, and another report an effectively decoupled ER and MR [8]. This *x* = 0.18 material, as it is cooled, first enters a FMM phase (below about 170 K), then a FMI phase (below about 120 K) [8,44]. Neither our *x* = 0.1 or *x* = 0.2 samples show such a three-phase behaviour with temperature.

The *x* = 0.2 compound is seen to undergo a MIT at ∼190 K (Figs. 5, 6). Our data is similar to a previous measurement which show a peak at 224 K [45] and very similar to data that show the feature at 190 K [41,42]. Our data is in agreement with the phase diagram that shows a transition to a FMM at approximately 190 K [30,39]. This is the sample of the lowest value of *x* that shows a significant ER. We estimate the ER at 1 mA relative to 0.1 mA to be -20% at $T_p$. This data is remarkable in that there is negligible ER away from the resistance peak, but strong ER near it, as may be seen from Fig. 6.

Others have made studies of this compound excited by large current [18,34,46]. Metastable states [46] and a variety of other hysteretic and time-dependent effects have been observed [18,34]. This current regime is quite different to the one used in the experiments reported here.

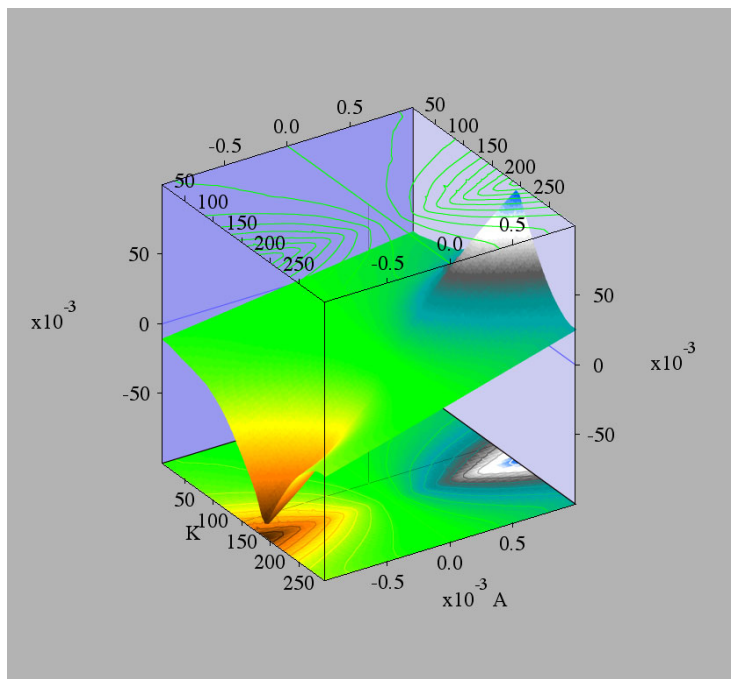

**Figure 5**
**La$_{0.8}$Ca$_{0.2}$MnO$_3$ voltage surface**. The voltage across the sample is shown on the vertical axis as a function of the applied current and the temperature.

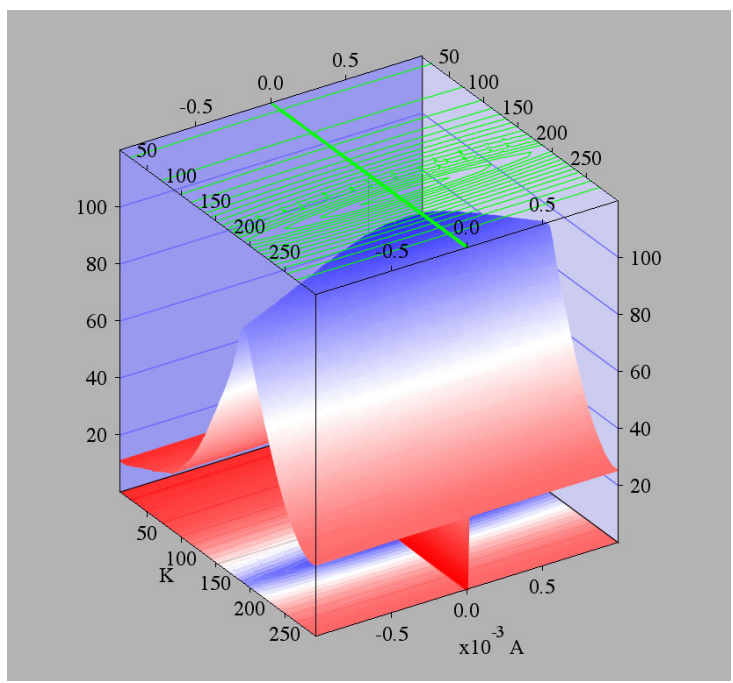

**Figure 6**
**La$_{0.8}$Ca$_{0.2}$MnO$_3$ resistance surface**. The resistance of the sample ($\Omega$) is shown on the vertical axis as a function of the applied current and the temperature.

### x = 0.3

This compound has been much studied and the resistance-temperature behaviour often reported [1,2,4,10,25,29,30,42,47]. The phase diagram predicts a MIT at ~250 K [30], and we observe $T_p$ = 245 K (Figs. 7, 8). This sample shows some ER at all temperatures, with the largest magnitude of ER around $T_p$. The ER at 1 mA relative to 0.1 mA is -60% at $T_p$. The overall behaviour in this bulk samples is similar to that to that reported previously in thin-film samples and is consistent with a PS percolation origin.

### x = 0.4

As seen in Figs. 9 and 10, this compound undergoes a broad MIT at ~175 K. To our knowledge data has not been given for this composition before. A report on the $x$ = 0.45 compound gives resistance-temperature data which is consistent with ours for the $x$ = 0.4 and $x$ = 0.5 compounds [42]. At $T_p$, the ER at 1 mA relative to 0.1 mA is about -10%, but reaches about -40% at 5 mA. To our knowledge, ER has not been reported previously in a sample of this composition.

### x = 0.5

This material is a charge-ordered insulator down to low temperatures [1].

Resistance-temperature curves previously reported [1,48] are somewhat different to our results shown in Figs. 11 and 12 which we attribute to grain size effects, as in previous measurements of polycrystalline material [49]. Our results are consistent with the data given in [1]. That work, in the FE configuration, has shown only a very small ER (< 1%) at the highest electric fields employed. Our data, Fig. 12, is the only known using an applied current. The ER is very small at room temperature. It increases with decreasing temperature, but seems to be maximum at a higher temperature than $T_p$.

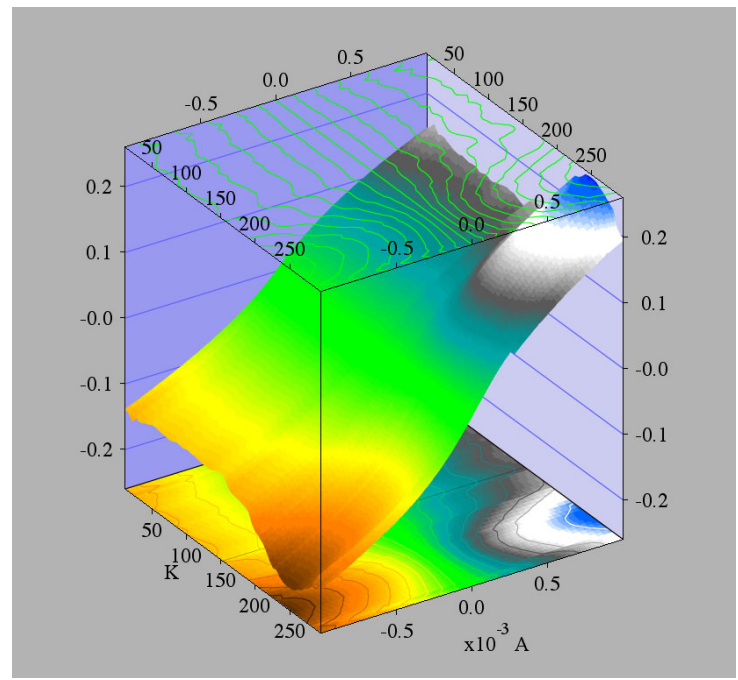

**Figure 7**
**La$_{0.7}$Ca$_{0.3}$MnO$_3$ voltage surface**. The voltage across the sample is shown on the vertical axis as a function of the applied current and the temperature.

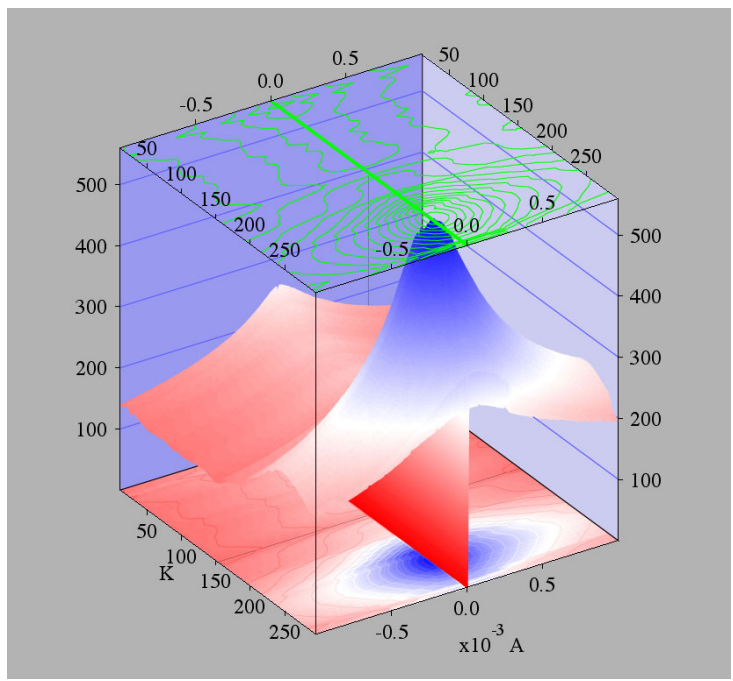

**Figure 8**
**La$_{0.7}$Ca$_{0.3}$MnO$_3$ resistance surface**. The resistance of the sample ($\Omega$) is shown on the vertical axis as a function of the applied current and the temperature.

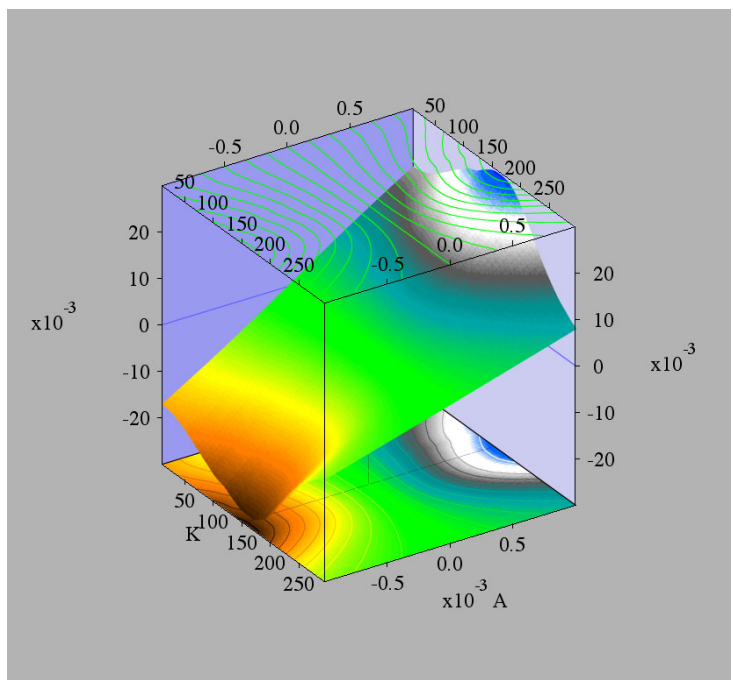

**Figure 9**
**La$_{0.6}$Ca$_{0.4}$MnO$_3$ voltage surface**. The voltage across the sample is shown on the vertical axis as a function of the applied current and the temperature.

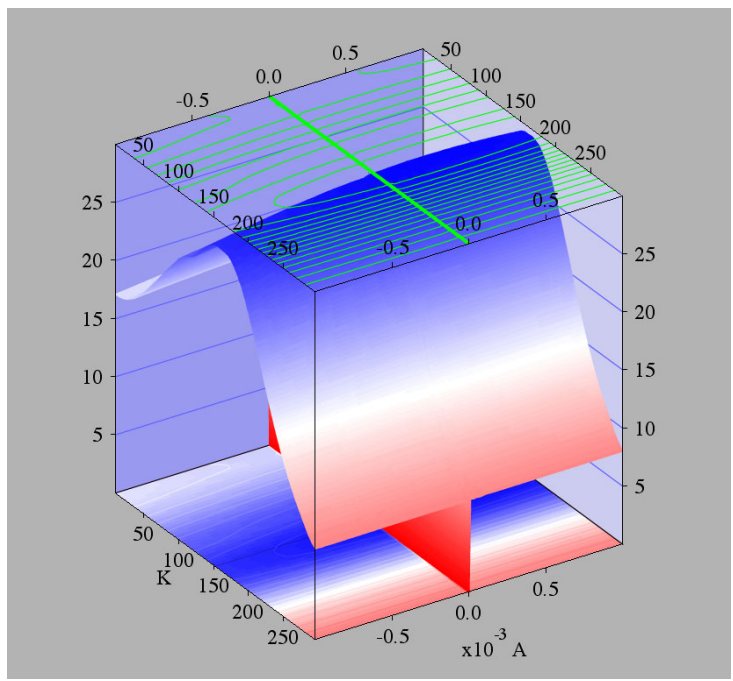

**Figure 10**
**La$_{0.6}$Ca$_{0.4}$MnO$_3$ resistance surface**. The resistance of the sample ($\Omega$) is shown on the vertical axis as a function of the applied current and the temperature.

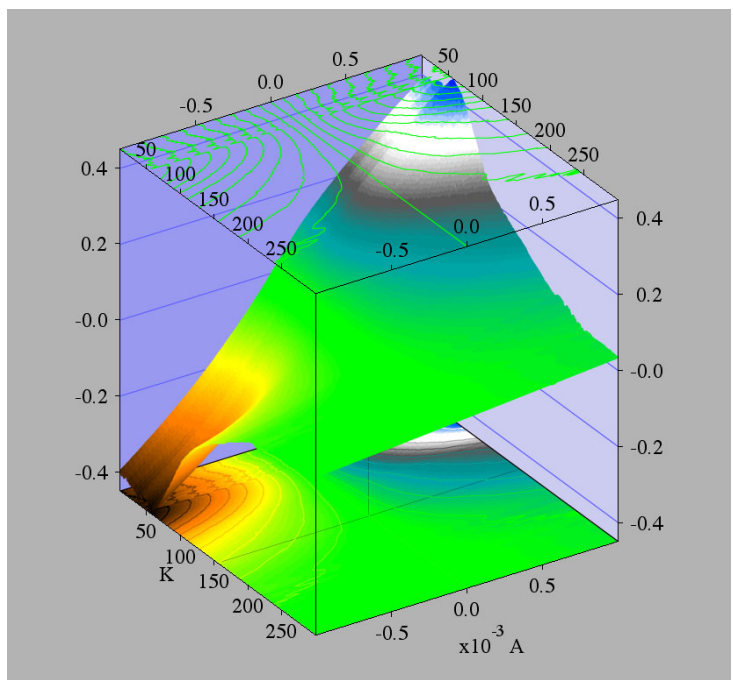

**Figure 11**
**La$_{0.5}$Ca$_{0.5}$MnO$_3$ voltage surface**. The voltage across the sample is shown on the vertical axis as a function of the applied current and the temperature.

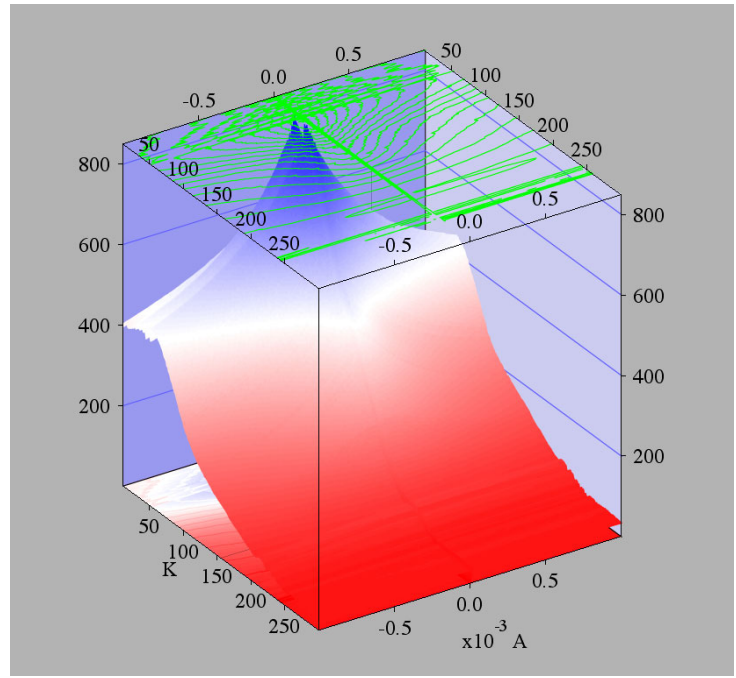

**Figure 12**
**La$_{0.5}$Ca$_{0.5}$MnO$_3$ resistance surface**. The resistance of the sample ($\Omega$) is shown on the vertical axis as a function of the applied current and the temperature.

### Joule heating

We are confident joule heating is not affecting our results. For a start, many other measurements which employ thin films in which large current densities are common, up to $1.5 \times 10^{10}$ A/m$^2$ [25]. We employ bulk samples, with considerably smaller current densities, estimated not to exceed 1 A/m$^2$. The bulk samples we employ also have much larger heat capacities than thin film samples. Second, Zhao *et al.* [4] have carefully studied the signature of heating as shown in R-T curves. In comparing their Fig. 1 (negligible joule heating) and their Fig. 2 (severe joule heating) all of our data is of the nature of the former. There is no shift in $T_p$ and no hysteresis. Third, joule heating would have the effect of increasing the ER above $T_p$ and decreasing the ER below $T_p$. We do not observe this, again supporting the assertion that joule heating is negligible in the experiments.

## Conclusion

Our results give a comprehensive picture of resistance as a function of both current and temperature in an important series of compounds La$_{1-x}$Ca$_x$MnO$_3$ ($x$ = 0–0.5). They illustrate a further aspect of the rich variety of physical phenomena in this family. ER is seen to be small in the compounds with $x$ = 0, 0.1, and 0.2. It becomes pronounced in the compounds $x$ = 0.3 and 0.4. It is generally strongest around $T_p$ but – especially in the 0.3 compound – is observed at all temperatures. Taken as a whole, the present data are consistent with the explanation of ER being due to PS. According to that picture, in materials where there is no phase coexistence, only a single insulating state, ER is small. Our result confirm that this explanation, first postulated for thin films [1,5], also holds in the bulk.

## Acknowledgements
We thank X. L. Wang and F. Gao for provision of samples. This work was supported by the Australian Research Council.